\documentclass[conference]{IEEEtran}
\IEEEoverridecommandlockouts

\usepackage{cite}
\usepackage{amsmath,amssymb,amsfonts}
\usepackage{algorithmic}
\usepackage{graphicx}
\usepackage{textcomp}
\usepackage{xcolor}
\def\BibTeX{{\rm B\kern-.05em{\sc i\kern-.025em b}\kern-.08em
    T\kern-.1667em\lower.7ex\hbox{E}\kern-.125emX}}

\usepackage{tikz}
\usepackage{caption}
\usepackage{blindtext}
\usepackage{tcolorbox}
\usepackage[final]{pdfpages}
\usepackage{lipsum,multicol}
\usepackage{xcolor}
\usepackage{tikz}
\usepackage{listings}
\usepackage{enumitem}
\usepackage{hyperref}
\usepackage{amsfonts}
\usepackage{wrapfig}
\usepackage{subcaption} 
\usepackage{adjustbox}
\usepackage{colortbl}
\usepackage{fancybox}
\usepackage{multirow}
\usepackage[normalem]{ulem}
\useunder{\uline}{\ul}{}
\usepackage{enumitem}
\usepackage{amsmath}
\usepackage{booktabs} 
\usepackage{array}
\usetikzlibrary{shapes.geometric, arrows.meta, positioning, fit}
\usepackage{balance}

\newtcolorbox{boxK}{
    fontupper = \small,
    sharpish corners, 
    boxrule = 0pt,
    toprule = 0pt, 
}

\newcommand{\secref}[1]{Sec.~\ref{#1}\xspace}
\newcommand{\figref}[1]{Fig.~\ref{#1}\xspace}

\definecolor{gradientplum}{RGB}{196,115,156}

\definecolor{gradientplum}{RGB}{196,115,156}

\newcommand{\ates}{\textit{ATEs}\xspace}

\newcommand{\approptoinn}[2]{\mathrel{\vcenter{
  \offinterlineskip\halign{\hfil$##$\cr
    #1\propto\cr\noalign{\kern2pt}#1\sim\cr\noalign{\kern-2pt}}}}}

\definecolor{gradientplum}{RGB}{196,115,156}
\newcommand\repository[1]{\textcolor{gradientplum}{\href{#1}{repository}}}

\begin{document}

\title{On Automated and Explainable Provenance of AI-Generated Code
}

\author{\IEEEauthorblockN{Alejandro Velasco, Nathan Wintersgill, Trevor Stalnaker, Oscar Chaparro, Denys Poshyvanyk}
\IEEEauthorblockA{\textit{School of Computing, Data Sciences and Physics} \\
\textit{William \& Mary} -
Williamsburg, VA, USA \\
\{svelascodimate, njwintersgill, twstalnaker, oscarch, dposhyvanyk\}@wm.edu}
}

\maketitle

\begin{abstract}

Generative AI for code generation has transformed software development, but it has also introduced a critical transparency problem: the origins of AI-generated code are opaque to the developers who use it, the organizations that deploy it, and the compliance professionals responsible for ensuring its legal and quality standards. Existing mitigations flag problematic outputs after the fact without explaining why a model produced them or how future generation could be improved. We present a research vision, grounded in a U.S. NSF-funded research grant, that argues that the next generation of CodeGenAI tools must be built on a foundation of explainable provenance: automated, post-hoc traceability that links generated code back to the prompt components, training data instances, global data features, and internal model components that caused its generation. We grounded this vision in empirical evidence from studies of software developers, model users, and compliance/legal professionals, which show that provenance information is a practical necessity that current tools do not provide. We characterize the problem across four traceability dimensions, outline a research program combining large-scale empirical studies with post-hoc causal and interpretability techniques, and identify the key open challenges that the community must address to realize this vision.
\end{abstract}


\section{Introduction}
\label{sec:introduction}

Generative AI for code generation (CodeGenAI) has fundamentally transformed how developers create software. Tools such as Claude Code, GitHub Copilot, Tabnine, and ChatGPT allow practitioners to generate complex functionality from natural-language prompts, dramatically accelerating development and lowering barriers to entry for non-expert programmers \cite{birdcopilot, liang2024survey}. As developers increasingly delegate implementation decisions to AI models, the origins of the resulting code become opaque to the developers themselves, to their organizations, and to the compliance professionals responsible for ensuring legal and quality standards.

As developers increasingly rely on CodeGenAI, the benefits of this technology do not come without costs. Numerous studies and real-life cases show that CodeGenAI can introduce significant security and legal risks, many of which can lead to financial, reputational, and legal consequences for organizations and society \cite{kahveci2023, pearce2022asleep, siddiq2022smells, yang2024memorization}. Specifically, CodeGenAI models have been shown to reproduce verbatim fragments of copyrighted code \cite{kahveci2023}, expose sensitive information memorized from training data \cite{yang2024memorization}, and generate insecure or low-quality code at scale \cite{pearce2022asleep, siddiq2022smells}. The ongoing lawsuit \textit{Doe v.\ GitHub}~\cite{doe2022github} has placed the legal status of AI-generated code under direct judicial scrutiny, and regulatory mandates 
now require Software Bills of Materials (SBOMs) to document the provenance of software components, including those produced with AI assistance: the US Executive Order on Cybersecurity~\cite{eo14028} mandated SBOMs for federal software procurement, and the EU Cyber Resilience Act \cite{cra2024} makes SBOMs a legally binding requirement for all products with digital elements placed on the market.

Tracing the origins of AI-generated code is an extraordinarily difficult problem. Three compounding factors contribute to this difficulty. First, the \emph{CodeGenAI supply chain} is long and opaque, spanning dataset curation, model pre-training, fine-tuning, deployment, and prompting, with each stage performed by different actors and each capable of influencing model output in ways that are difficult to disentangle \cite{lee2023supply}. It also often contains deficiencies in documentation, such as missing references to necessary components like datasets~\cite{stalnaker2025empirical}. Second, the \emph{black-box nature} of large language models (LLMs) makes it difficult to inspect why a particular prompt yields a particular output, and understanding model behavior requires interpretability techniques that are still maturing \cite{doshivelez2017rigorous}. Third, \emph{training data} are often inaccessible, withheld for legal or commercial reasons, making it impossible to directly verify whether a generated snippet originates from a specific source~\cite{longpre2024}.
\looseness=-1

Existing mitigation strategies address this problem only partially. Filtering layers and specialized detection tools such as CopyrightCatcher flag problematic output after generation, without explaining \emph{why} a model produced them or informing how future generation could be improved \cite{patel2024copilot}. Attribution-by-link approaches, in which AI assistants cite web sources along with generated code, have been shown to produce largely irrelevant references \cite{bifolco2025llms}. No existing solution provides developers with an actionable \emph{and explainable} traceability between AI-generated code and the factors that caused its generation.

This paper presents a research vision grounded in  research grant IIS-2533367~\cite{nsf_grant} funded by the U.S. National Science Foundation (NSF). We argue that the next generation of CodeGenAI tools must be built on a foundation of \emph{explainable provenance}: automated traceability that links generated code back to its contributing causes, including prompt components, training data instances and their global characteristics, and internal model components, presented in a form that enables developers and compliance practitioners to make informed decisions about the code they integrate. This vision requires advances along two fronts: (i) a deeper empirical understanding of what provenance information different stakeholders actually need and how they currently seek it, and (ii) new post-hoc techniques that provide explainable traceability without requiring model retraining.

\section{The Need for Explainable Provenance}
\label{sec:background}

Understanding the origins of AI-generated code has become a practical necessity for three groups of stakeholders who occupy distinct positions in the CodeGenAI supply chain: \emph{model developers}, who curate training data and build and fine-tune models; \emph{model users}, who integrate AI-generated code into software projects; and \emph{compliance practitioners}, who audit AI-generated artifacts against legal and organizational requirements \cite{lee2023supply, stalnaker2024sboms}.

\textbf{Evidence from practitioners.}
Our past survey of 574 software developers worldwide found that while $89.7\%$ regularly use CodeGenAI tools, $70.1\%$ lack or do not have processes to document their use, creating significant traceability gaps~\cite{stalnaker2024licensing}. Follow-up interviews revealed that developers resort to manual searches on GitHub and Stack Overflow to verify the origins of AI-generated code, a practice that is both inefficient and unreliable at scale. We conducted another survey of 138 software practitioners across five stakeholder groups. The survey established that tracking software provenance is a critical concern, particularly for AI systems where missing lineage information can obscure issues such as model poisoning, training data bias, or inadvertent inclusion of harmful content~\cite{stalnaker2024sboms}. The legal professionals surveyed in a complementary study reinforced this picture~\cite{wintersgill2024legal}: we found that provenance information is a prerequisite for maintaining license compliance, and existing tools such as Black Duck and FOSSology do not offer coverage of AI-generated artifacts.

\textbf{What provenance must explain.}
The information needs of these stakeholders converge on four traceability dimensions. \emph{Prompt-to-output traceability} concerns which components of the prompt contribute to the generation of non-compliant or low-quality code. \emph{Instance-to-output traceability} concerns which training data or external code instances are associated with a given output. \emph{Factor-to-output traceability} concerns the global characteristics of the training data, such as the prevalence of restrictively licensed code or specific syntactic constructs, that lead the model to learn patterns producing problematic outputs. Finally, \emph{model component-to-output traceability} concerns the internal components of the model, such as specific neuron circuits or attention heads, responsible for a given generation. Together, these four dimensions define the scope of explainable provenance as we envision it.

\textbf{Why current approaches fall short.}
Provenance methods developed for traditional software, including  clone detection, hashing, and software bertillonage \cite{davies2013bertillonage, german2009siblings}, were designed for settings where copying produces an identifiable artifact. In the generative setting, the relationship between a model output and any contributing source is probabilistic, mediated by billions of learned parameters, and shaped by prompt context that shifts with every query. Intrinsic source attribution, which embeds citation information during training, requires access to the training pipeline and remains prone to citation hallucinations \cite{khalifa2024source}. Post-hoc retrieval approaches can surface similarities between generated and existing code, but stop short of explaining the causal role of training data characteristics or model components in producing a given output \cite{majdinasab2024trained}. The result is a significant gap: stakeholders have no principled, automated way to understand the causes behind a specific AI-generated output, nor guidance on how to adapt prompts, curate training data, or refine models to prevent problematic generations in the future.
\looseness=-1

\begin{figure*}
	\centering
	\includegraphics[width=\linewidth]{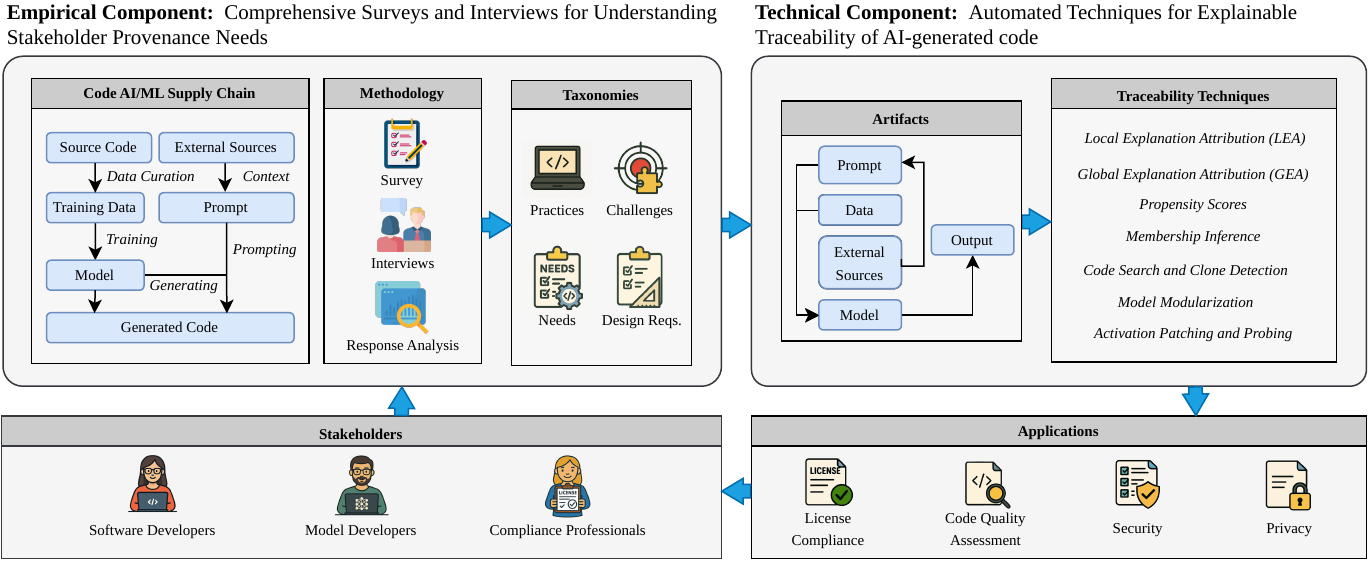}
	\caption{Overview of the proposed research vision.}
	\label{fig:research-program}
\end{figure*}

\section{A Research Vision for Explainable Provenance}
\label{sec:approach}

We envision a research program organized around two interconnected goals. The first is empirical: to develop a comprehensive understanding of how stakeholders currently engage with provenance during the development and use of CodeGenAI systems, and what information they need to do so effectively. The second is technical: designing automated, post-hoc techniques that provide explainable traceability between AI-generated code and the artifacts and factors that caused its generation. \figref{fig:research-program} illustrates the overall research vision.

\subsection{Comprehensive Surveys and Interviews for Understanding
Stakeholder Provenance Needs}
\label{sec:vision:empirical}

Despite the growing adoption of CodeGenAI tools, little is known about how model developers, model users, and compliance practitioners engage with provenance in practice. \textit{What information do they seek when evaluating AI-generated code?} \textit{What challenges prevent them from obtaining it?} \textit{What would effective tool support look like from their perspective?} These questions remain largely unanswered, and their answers are a prerequisite for designing provenance techniques that are both technically sound and practically useful.

To address this gap, we propose a large-scale empirical study that 
combines an online survey and in-depth follow-up interviews with the 
three stakeholder groups. Each group plays a distinct role in the AI 
supply chain. First, \emph{model developers} curate datasets and train, 
fine-tune, and deploy CodeGenAI models. Second, \emph{model users} 
integrate AI-generated code into software projects via prompting. 
Third, \emph{compliance practitioners}, including legal professionals 
and Open Source Program Office (OSPO) staff, evaluate AI-generated 
artifacts against legal and organizational mandates~\cite{wintersgill2024legal}.
\looseness=-1

The study will be conducted iteratively over two years, allowing findings from early rounds to sharpen the questions posed in later ones. Survey questionnaires will be designed to uncover practices, needs, and challenges in provenance across the AI supply chain lifecycle, covering topics such as how stakeholders document the sources of training data, what metadata they need about AI-generated outputs to reason about quality and compliance, what traceability mechanisms would make their workflows more efficient, and what challenges they face in tracing low-quality or non-compliant code back to its origins. Respondents who provide insightful answers will be invited to in-depth follow-up interviews of 40--60 minutes to collect richer, contextualized information about their specific needs and practices.

Responses will be analyzed qualitatively using open coding and card sorting to develop structured taxonomies \cite{stalnaker2024sboms, stalnaker2024licensing}. These taxonomies will capture provenance practices, needs, challenges, and design requirements for each stakeholder group, organized around the stages of the AI supply chain and the artifact types involved, from raw training data and model checkpoints to prompts and generated code. The design requirements will inform the development of the envisioned techniques described in \secref{sec:vision:technical}.  Inter-rater reliability will be measured throughout, and taxonomy development will proceed until theoretical saturation is reached. The resulting taxonomies will define the functional and non-functional requirements that the techniques described next must satisfy.
\looseness=-1

\subsection{Automated Techniques for Explainable Traceability}
\label{sec:vision:technical}

The technical core of the vision is a suite of post-hoc traceability techniques that address each of the four provenance dimensions introduced in \secref{sec:background}. 

\begin{figure*}[t]
    \centering
    \includegraphics[width=\linewidth]{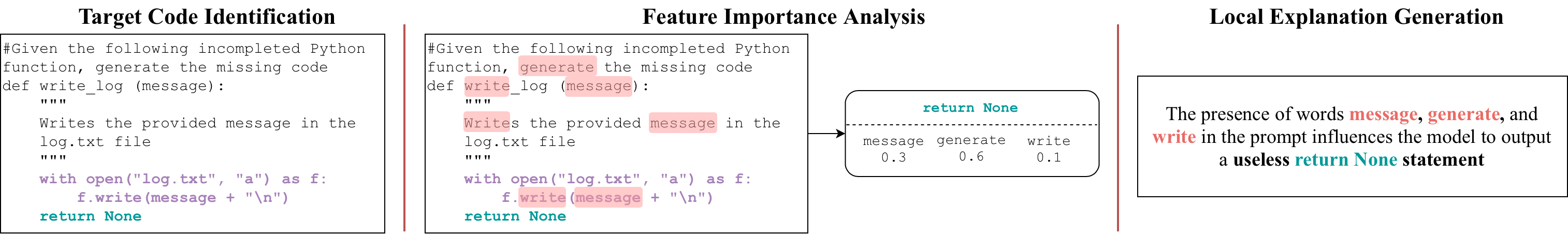}
    \caption{Workflow of Local Explanation Attribution (LEA). Given a
    prompt and a smelly generated fragment (highlighted in green),
    SHAP values identify the prompt tokens most responsible for the
    problematic output, and a natural-language explanation is
    generated for the developer.}
    \label{fig:lea}
\end{figure*}

\textbf{Prompt-to-output traceability.}
We propose two complementary techniques for linking prompts to
generated outputs. The first is a \textit{propensity scoring} approach that analyses token-level log-probabilities from the model's output distribution to estimate the likelihood that a given prompt will lead to non-compliant or low-quality code \cite{velasco2025propensity, velasco_causal}. The second is \textbf{L}ocal \textbf{E}xplanation \textbf{A}ttribution (\textit{LEA}), a post-hoc interpretability technique that identifies the prompt tokens most responsible for a problematic generation \cite{velasco2025propensity, velasco_causal, yang2025sectmi, yang2025privacy}, using feature attribution methods such as SHAP \cite{lundberg2017shap} or rationale-based approaches \cite{khati_enabling_2026, vafa2021rationales}. As illustrated in \figref{fig:lea}, LEA proceeds in three steps: identifying the generated code fragment that is non-compliant or low-quality, applying feature attribution to rank the prompt tokens by their influence on that fragment, and generating a natural-language explanation that model users can act on to refine their prompts.

\textbf{Instance-to-output traceability.}
Linking generated code to specific training data or external code instances requires a retrieval pipeline that combines traditional clone detection, neural similarity models, and membership inference attacks \cite{hu2022membership, yang2025sectmi}. The pipeline adapts to three scenarios depending on data availability. When the training set is accessible, similarity-based retrieval can be applied directly against indexed training instances. When it is not, web-scale search using code features extracted from the generated output serves as a proxy. When the model provides source links, those links are crawled, parsed, and verified through clone detection to assess their relevance and associated licensing terms \cite{bifolco2025llms}. In all scenarios, causal analysis of training dynamics can further reveal which artifact types carry the highest leakage risk, informing where retrieval efforts should be concentrated \cite{yang2025privacy}.

\begin{figure}[t]
    \centering
    \includegraphics[width=0.8\linewidth]{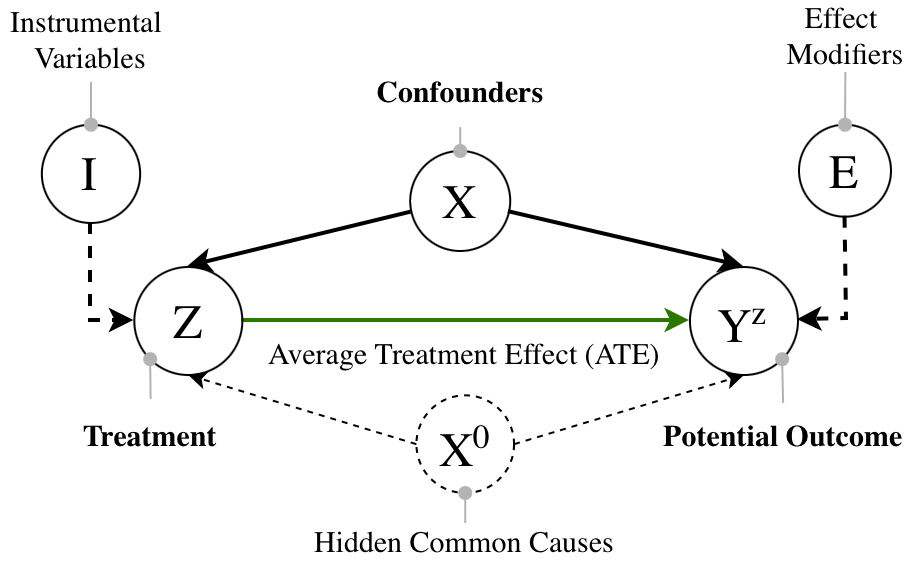}
    \caption{Structure of the causal graphs used by \textit{GEA}. Treatments $Z$ represent training data factors; confounders $X$ account for contextual variables such as model architecture and data distribution; the green arrow represents the Average Treatment Effect (ATE) estimated via do-calculus.}
    \label{fig:gea}
\end{figure}

\textbf{Factor-to-output traceability.}
Understanding how global characteristics of the training data influence model behavior requires moving beyond instance-level retrieval to causal analysis. We propose \textbf{G}lobal \textbf{E}xplanation \textbf{A}ttribution (\textit{GEA}), a framework that extends causal inference techniques developed for code models \cite{palacio2024docode} to the provenance setting. As shown in \figref{fig:gea}, \textit{GEA} uses causal graphs to model the relationships between training data factors (treatments $Z$), such as code complexity, syntactic constructs, and license distributions, and the propensity of the model to generate problematic outputs (potential outcomes $Y^Z$), while accounting for confounders $X$, instrumental variables $I$, hidden common causes $X^0$, and effect modifiers $E$. By estimating average treatment effects (\ates) through do-calculus and standard adjustment techniques, \textit{GEA} produces actionable, interpretable insights into how training data composition shapes model behavior, without requiring model retraining.

\textbf{Model component-to-output traceability.}
At the finest level of granularity, provenance requires linking generated outputs to specific internal components of the model, which is possible only for open-weight models. We propose to investigate three complementary approaches: functional model modularization, which identifies sub-networks associated with specific behaviors \cite{imtiaz2023decomposing}; sparse autoencoders, which extract interpretable low-dimensional representations from internal activations \cite{bricken2023monosemanticity}; and activation patching combined with probing, which tests the effect of modifying specific neuron circuits on model outputs and identifies which internal representations encode patterns associated with non-compliant or low-quality code \cite{meng2023locating}.
\looseness=-1
\section{Challenges and Open Questions}
\label{sec:challenges}

Realizing the vision of an explainable provenance for AI-generated code requires addressing several fundamental challenges that cross the empirical and technical dimensions of the research program.

\textbf{Defining provenance granularity.}
Provenance information can be provided at many levels of granularity, from individual tokens to entire functions, from specific training instances to broad data characteristics, from single attention heads to functional model modules. The appropriate granularity depends on the stakeholder and the task at hand. A model developer diagnosing systematic quality issues needs different information from a compliance practitioner auditing a specific code snippet for license violations. Determining which granularities are most actionable for each stakeholder group and designing techniques that can operate across multiple levels of granularity without prohibitive computational cost remain an open problem that the empirical component of the research program must inform.

\textbf{Scalability under training data inaccessibility.}
Many of the most widely used CodeGenAI models are proprietary and their training data is withheld for legal or commercial reasons. Provenance techniques that assume access to training data are therefore inapplicable to a large and growing fraction of real-world deployment scenarios. Membership inference attacks offer a partial solution, but their accuracy is reduced for large models with diverse training corpora, and provide statistical evidence rather than definitive attribution \cite{hu2022membership}. Developing provenance techniques that remain informative and reliable in black-box or grey-box settings, where only model outputs or output probabilities are accessible, represents one of the most pressing open challenges in this space.

\textbf{Causal attribution in the presence of confounders.}
Establishing causal links between training data characteristics and model output is complicated by the presence of confounders that are difficult to observe or control. Model architecture, training hyperparameters, tokenization strategies, and data preprocessing decisions all influence model behavior in ways that interact with training data factors in complex, non-linear ways. Causal graphs constructed from domain knowledge may miss important confounders, leading to biased estimates of treatment effects \cite{palacio2024docode}. Developing robust causal discovery procedures that can identify relevant confounders from data and sensitivity analyses that quantify the impact of unobserved confounding on provenance conclusions will be essential for the credibility of the \textit{GEA} framework.

\textbf{Evaluating provenance without ground truth.}
Provenance techniques require evaluation, but ground truth is difficult to obtain in realistic settings. For commercial models, the true training data is unknown. For open-source models, the relationship between a specific generated output and any particular training instance is probabilistic rather than deterministic. Constructing evaluation benchmarks that simulate realistic provenance scenarios, with controlled license violations and known causal relationships between training data and output, requires careful design to avoid artificial simplifications that do not reflect the complexity of real-world code generation~\cite{velasco2025propensity}. Human evaluation is an alternative, but inter-annotator agreement on provenance judgments, particularly for license compliance, is itself a research question given the inherent ambiguity of copyright law \cite{wintersgill2024legal}.

\textbf{Bridging technical provenance and legal reasoning.}
Even technically precise provenance information may not translate directly into legally actionable conclusions. The copyright law does not operate on similarity scores or membership inference probabilities; it requires judgments about substantial similarity, the expression of ideas rather than the ideas themselves, and the intent of the parties involved \cite{kahveci2023}. Designing provenance tools that present technical evidence in forms that are interpretable and useful to legal professionals, without overstating or understating their significance, requires close collaboration between software engineering researchers and legal scholars. This interdisciplinary challenge may ultimately prove as difficult as the technical problems the tools are designed to solve.
\section{Related Work}
\label{sec:related_work}

Our vision is based on four areas: empirical studies of AI-assisted
development, software provenance and compliance, source attribution
for generated code, and interpretability of code models. Each
contributes motivation or technical building blocks, but none provides end-to-end, post-hoc traceability across all four provenance dimensions we identify.

\textbf{AI-assisted programming and risks.}
Studies of GitHub Copilot and related tools show that developers use CodeGenAI for completion, testing, debugging, and review \cite{ liang2024survey}, but do not consistently document the presence or origin of AI-generated code. Previous work establishes that generated code can introduce security vulnerabilities \cite{pearce2022asleep}, code smells \cite{siddiq2022smells, velasco2025propensity}, and memorized training fragments \cite{yang2024memorization}. Detecting problematic snippets is insufficient: developers and compliance practitioners also need to understand which prompt elements, training examples, and model components contributed to them.

\textbf{Provenance, SBOMs, and compliance.}
Clone detection, dependency analysis, and software bertillonage \cite{german2009siblings, davies2013bertillonage} support provenance recovery for traditional software. Work on Software Bills of Materials shows that stakeholders need structured lineage across the supply chain \cite{stalnaker2024sboms}. Legal and empirical studies further show that practitioners need provenance evidence connected to legal concepts such as attribution and substantial similarity \cite{kahveci2023, lee2023supply, stalnaker2024licensing, wintersgill2024legal}. These approaches assume provenance is attached to identifiable artifacts. CodeGenAI breaks this assumption: a generated function may reflect many training instances and prompt choices without being a direct clone of any single source.

\textbf{Source attribution and code inclusion.}
Source-aware training teaches models to associate outputs with source identifiers \cite{khalifa2024source}, but requires access to the training pipeline. Post-hoc approaches are therefore necessary. AI assistants that provide source links have been shown to produce largely irrelevant references \cite{bifolco2025llms}. Model-agnostic techniques can detect whether a code artifact was included in a training set \cite{majdinasab2024trained}, and membership inference attacks provide complementary statistical evidence \cite{hu2022membership}. Recent work further shows that semantically equivalent code transformations can be used to probe the robustness of membership inference for code models \cite{yang2025sectmi}, while causal analysis of training dynamics reveals that leakage risks differ substantially across personally identifiable information types and correlate with how easily each type is learned \cite{yang2025privacy}. These works address instance-to-output traceability but do not explain how prompt choices, corpus-level factors, or model internals caused a given output.

\textbf{Interpretability for code models.}
Feature-attribution and rationale-based methods identify which prompt tokens most influence a model prediction \cite{lundberg2017shap, vafa2021rationales}. Causal frameworks for code models distinguish correlation from causal influence \cite{palacio2024docode}, motivating our GEA direction. Recent work applies causal analysis to measure, explain, and mitigate smell propensity in LLM-generated code, showing that prompt formulation and model architecture causally shape structural quality \cite{velasco_causal}. At the model-internal level, sparse autoencoders and activation patching can localize learned behaviors to functional modules and internal circuits \cite{bricken2023monosemanticity, meng2023locating}, providing a foundation for model component-to-output traceability. These techniques have not yet been connected to provenance tasks such as license compliance or code quality audit.

Existing approaches address individual levels of provenance evidence in isolation. Our agenda treats provenance as a multi-level explanation problem spanning all four traceability dimensions, operating post-hoc without access to proprietary training pipelines.
\section{Conclusions \& Future Plans}
\label{sec:conclusion}


We have presented a vision for explainable provenance of AI-generated 
code, framed as a multi-level traceability problem spanning prompt 
components, training data instances, global data features, and internal 
model components. Empirical evidence shows that current practices leave 
significant provenance gaps for all three stakeholder groups in the 
CodeGenAI supply chain, and this vision motivates a research program 
along two complementary fronts. The empirical front will produce 
taxonomies of stakeholder provenance needs, challenges, and design 
requirements through large-scale surveys and interviews, while the 
technical front will develop and evaluate post-hoc traceability 
techniques across all four provenance dimensions. Several open 
challenges remain (\secref{sec:challenges}), from determining 
actionable provenance granularities to bridging the gap between 
technical evidence and legal reasoning. We invite the community to 
engage with this vision and contribute to defining the methods and 
evaluation standards that explainable provenance will require.

\section{Acknowledgments}
\label{sec:acknowledgments}

This work has been funded by the U.S. NSF award
IIS-2533367~\cite{nsf_grant}. We would like to thank Massimiliano Di Penta from the University of Sannio for his valuable contributions to the grant proposal. The opinions, findings and conclusions
expressed in this work are solely those of the authors and do
not necessarily reflect the views of the sponsors.

\balance
\bibliographystyle{ieeetr}
\bibliography{utils/references}

\end{document}